\documentclass[amsmath, amsthm, amssymb, floatfix, reprint, onecolumn, notitlepage, nofootinbib, 12pt]{revtex4-1}
\usepackage{setspace}
\usepackage{amsmath}
\usepackage{breqn}
\usepackage{graphicx}
\usepackage[nearskip,margin = 0pt,caption=false]{subfig}

\usepackage{verbatim}
\usepackage{amsfonts}
\usepackage{amssymb}
\usepackage{epstopdf}
\usepackage{lipsum}
\usepackage{siunitx}
\usepackage{xcolor}
\usepackage{array}
\usepackage{braket}
\usepackage[normalem]{ulem}
\usepackage[resetlabels,labeled]{multibib}
\DeclareGraphicsExtensions{.pdf,.eps,.png,.jpg,.mps}
\usepackage{etoolbox}
\usepackage{subfiles} 
\renewcommand{\baselinestretch}{2.0}

\newcommand{\hl}[1]{{\color{black} #1}}

\begin{document}
\title{
Platform-agnostic waveguide integration of high-speed photodetectors with evaporated tellurium thin films}
\baselineskip=12pt
\author{Geun Ho Ahn$^{1, \dagger, *}$, Alexander D. White$^{1, \dagger, *}$, Hyungjin Kim$^{2, 3, \dagger}$, Naoki Higashitarumizu$^{2, 3}$, Felix M. Mayor$^{1}$, Jason F. Herrmann$^{1}$, Wentao Jiang$^{1}$, Kevin K. S. Multani$^{1}$, Amir H. Safavi-Naeini$^{1}$, Ali Javey$^{2, 3}$, Jelena Vu\v{c}kovi\'{c}$^{1}$\\
\vspace{+0.05 in}
$^1$E. L. Ginzton Laboratory, Stanford University, Stanford, CA 94305, USA.\\
$^2$Department of Electrical Engineering and Computer Sciences, University of California, Berkeley, CA, USA.\\
$^3$Materials Sciences Division, Lawrence Berkeley National Laboratory, Berkeley, CA, USA.\\
{\small $^{\dagger}$These authors contributed equally to this work.}\\
{\small $*$ gahn@stanford.edu, adwhite@stanford.edu}}
\baselineskip=24pt

\begin{abstract}
    \renewcommand{\baselinestretch}{2.0}
    \noindent \textbf{Many attractive photonics platforms still lack integrated photodetectors due to inherent material incompatibilities and lack of process scalability, preventing their widespread deployment. 
    Here we address the problem of scalably integrating photodetectors in a photonic platform-independent manner. Using a thermal evaporation and deposition technique developed for nanoelectronics, we show that tellurium (Te), a quasi-2D semi-conductive element, can be evaporated at low temperature directly onto photonic chips to form air-stable, high-responsivity, high-speed, ultrawide-band photodetectors. 
    \hl{We demonstrate detection at visible, telecom, and mid-infrared wavelengths, a bandwidth of more than \hl{40 GHz}, and platform-independent scalable integration with photonic structures in silicon, silicon nitride and lithium niobate}.}
\end{abstract}
\maketitle
\renewcommand{\baselinestretch}{2.0}

\newpage

Upcoming applications in nonlinear and quantum photonics have led to the proliferation of alternative material platforms that offer wider transparency windows, lower loss, and a diversity of optical nonlinearities \cite{okawachi2011octave,jung2021tantala, mishra2021mid,lukin20204h,dory2019inverse, bartholomew2020chip}. \hl{Applications of these platforms include mid-IR circuits for environment sensing and communication, nonlinear optics for frequency conversion and generation and metrology, and ultra-high-speed data encoding.} Integrated photodetectors play a critical role in these technologies, serving as the primary readout for many systems \cite{Atabaki:2018:Nature, pointcloud2021}. However, a vast majority of novel photonic material platforms do not have easily integratable photodetectors, and instread rely on external detection. This lack of on-chip photodetection prevents the scaling and deployment of many photonic integrated circuits (PIC) systems.

Current approaches to on-chip photodetection are either platform dependent or challenging to scale.
For instance, traditional photodiode materials such as germanium and III-V require stringent material synthesis processing like high synthesis temperature and crystal lattice matching to their host material \cite{michel2010high, liang2010hybrid, beling2013inp}, and are therefore limited to operation on silicon on insulator (SOI) or III-V platforms. \hl{Recent work depositing these materials in amorphous form is a promising avenue towards platform independence, but the reduced carrier mobility in amorphous material necessitates nano-scale fabrication features \cite{salamin2018100}.}
Two dimensional (2D) materials, many of which are photo-active \cite{Liu2016, Mak2016, youngblood2015waveguide, maiti2020strain, Flory2020, amani2018solution, tong2020stable}, have no such requirement of crystal lattice matching requirement owning to their naturally terminated surface, and can be transferred to nearly any substrate \cite{castellanos2014deterministic, marconi2021photo, liu2021silicon}. However, manufacturing, stability, and scaling remain as significant obstacles for the practicality of 2D photodetector integration \cite{long2019progress, chen2015environmental}. The mechanical exfoliation process used to make the 2D based photodetector is currently a stochastic, labor-intensive manual process, and while there are constant advancements in large area synthesis of 2D materials \cite{Kang2015}, there are still many outstanding challenges, including small grain size, strain, and material transfer \cite{long2019progress}.

Here, we show that thin film tellurium can be lithographically defined and evaporated at low temperature directly onto photonic structures to form easy-to-fabricate functional photodetectors. 
We utilize a thermal evaporation and deposition technique developed for nanoelectronics that forms crystalline quasi-2D structure with extremely high carrier mobility \cite{zhao2020evaporated, Zhao2021AM, tan2020evaporated}. \hl{The scalability of thermal evaporation and lithographic definition of active material allows us to demonstrate an array of 64 functional detectors fabricated in parallel.}
\hl{Additionally, the high mobility of the evaporated Te allows us to demonstrate state of the art device bandwidths of over 40 GHz}. These detectors can be fabricated on a variety of novel PIC platforms (\hl{here we show silicon, silicon nitride, and lithium niobate}), and the small bandgap of the evaporated tellurium allows operation from visible to mid infrared. Under low power and low frequency optical excitations, avalanche gain allows us to reach responsivities up to 1.9~AW$^{-1}$ at 520~nm, 0.8~AW$^{-1}$ at 1550~nm, and 0.15~AW$^{-1}$ at 2360~nm. We find that the evaporated tellurium is air stable, with no significant degradation in devices left in air for over a one month period, corroborating observations in flakes \cite{tong2020stable}. Finally, we show that because we can integrate these photodetectors directly on top of photonic structures, we can use integrated devices like micro-resonators to enhance the absorption of the detector and reduce relative dark current.

\begin{figure}[t]
\centering
\includegraphics[width=1\linewidth]{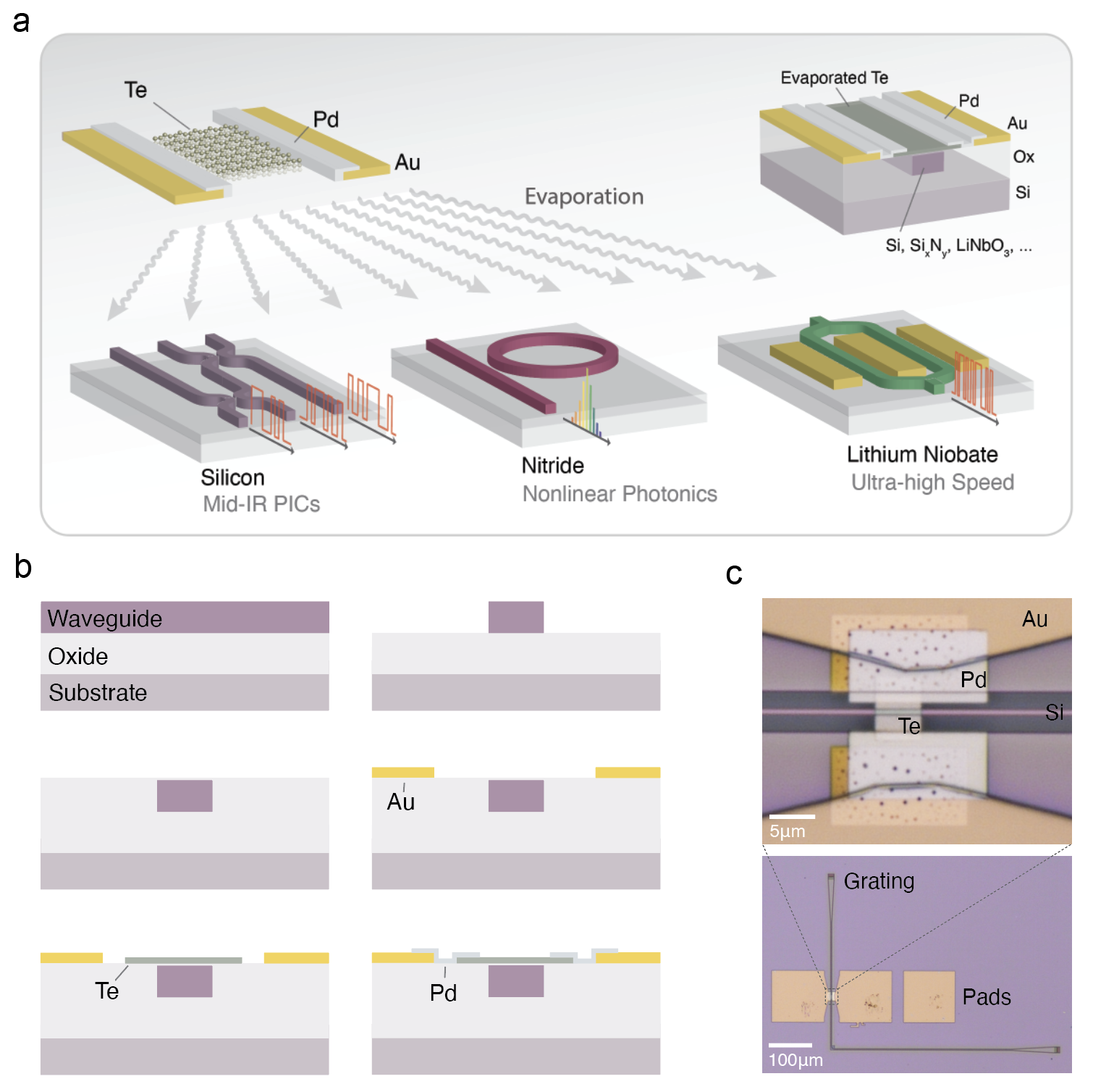}
\caption{\label{fig:Fig1}{\bf{Tellurium Photodetectors}} \textbf{(a)} \hl{Schematic of device.} \textbf{(b)} Fabrication process flow. SOI or any other photonics platform is etched to form waveguides. HSQ liquid glass is used to fill gaps and smooth the top surface. Gold pads, Tellurium devices, and finally Palladium contacts are lithographically defined and evaporated.  \textbf{(c)} Optical micrographs of a tellurium device (top) and its connection to waveguides and electronics (bottom). 
}
\end{figure}

\hl{Simple lithographic fabrication of Tellurium photodectors would allow for interfacing with photonics platforms that lack photodection including mid-IR silicon circuits, nonlinear photonics in silicon nitride, and ultra-high speed modulators in lithium niobate (FIG 1a).}
As shown schematically in FIG 1b, the fabrication process for the Te photodetectors is scalable and straightforward. We first define waveguide structures on an integrated photonics platform. To ensure even deposition onto the waveguide structure, we clad the waveguides with oxide; here we achieve this by spinning on and annealing Diluted H-SiOx (HSQ) liquid glass. We then lithographically define and deposit through evaporation of gold connections, the tellurium photoconductor thin film, and palladium (Pd) contacts, FIG 1a-c. The as-deposited Te thin films are poly-crystalline, as shown in Supplementary FIG S1, and optical images of patterned Te thin films after liftoff on PIC is shown in Supplementary FIG S2. The palladium contacts are chosen for \hl{relatively low contact resistance to the tellurium \cite{Peide_Ye_Contacts}, allowing the photodetector to operate in photoconductive mode}. However, other choices of metals such as asymmetric contacts such as Pd and  can allow for Schottky diode connections for lower dark currents \cite{Peide_Ye_Contacts}. By modulating the thickness of the tellurium layer, we can achieve a large range of optical bandgaps, allowing for detection from the visible to mid infrared \cite{zhao2020evaporated}. Though the tellurium layers used are extremely subwavelength, due to the proximity of the waveguide, nearly all of the light can be absorbed in just a few $\mu$m of propagation length, as shown by the length dependent transmission measurement of tellurium films shown in \hl{Supplementary FIG S3}. The combination of low temperature evaporation onto oxide and controllable bandgap allow for easy fabrication on top of nearly any integrated photonics platforms for use with any wavelength, without damaging photonics, metal, or active structures already fabricated.







\begin{figure}[b]
\centering
\includegraphics[width=1\linewidth]{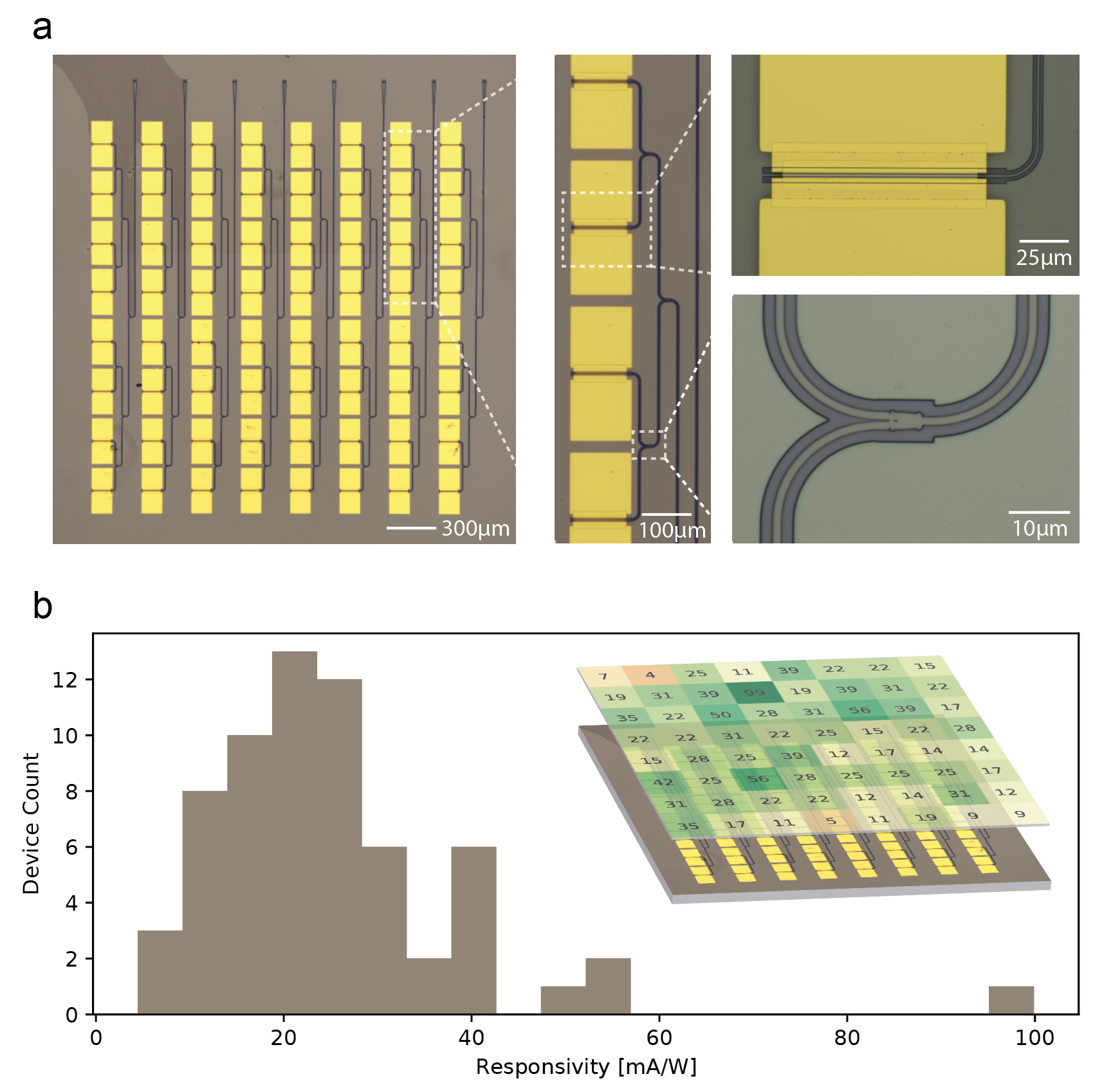}
\caption{\hl{ \label{fig:Fig1}{\bf{Detector Array}} \textbf{(a)} Optical micrographs of 8x8 device array of Te photodetectors on silicon nitride. Each column of devices is fed optically by a grating coupler and splitter tree. \textbf{(b)} Histogram and mapping of device responsivity measured with 2~$\mu$W of 1550~nm optical power at 10~MHz. }}
\end{figure}

\hl{To test the scalability and reliability of this integration of Te photodetectors scheme, we deposit in parallel a 64 element array of Te photodetectors on silicon nitride PIC, FIG 2a. Here, each column of 8 devices is fed by an underlying PIC that splits light incident on a grating coupler into the 8 photodetector waveguides. By sending modulated light through the gratings, we can measure the device responsivity of every device, as shown in FIG 2b. We find that every device in the array shows a measurable photo-response, with a distribution centered around 25 mAW$^{-1}$. 
Through transfer length measurement (TEM) of Te films, we find that the devices have a contact resistance of approximately 250~$\Omega$, as shown in Supplementalary FIG S4. As we measure the output with a 50~$\Omega$ load, the 25 mAW$^{-1}$ responsitivy corresponds to an internal quantum efficiency of 10\%.  
Additionally, by measuring the resistance, 27~k$\Omega$, and dark current, 114~$\mu A$,  of a representative Te photodetector device, we estimate the noise effective power (NEP) to be 27~pW/$\sqrt{\text{Hz}}$}.

\begin{figure}[b]
\centering
\includegraphics[width=1\linewidth]{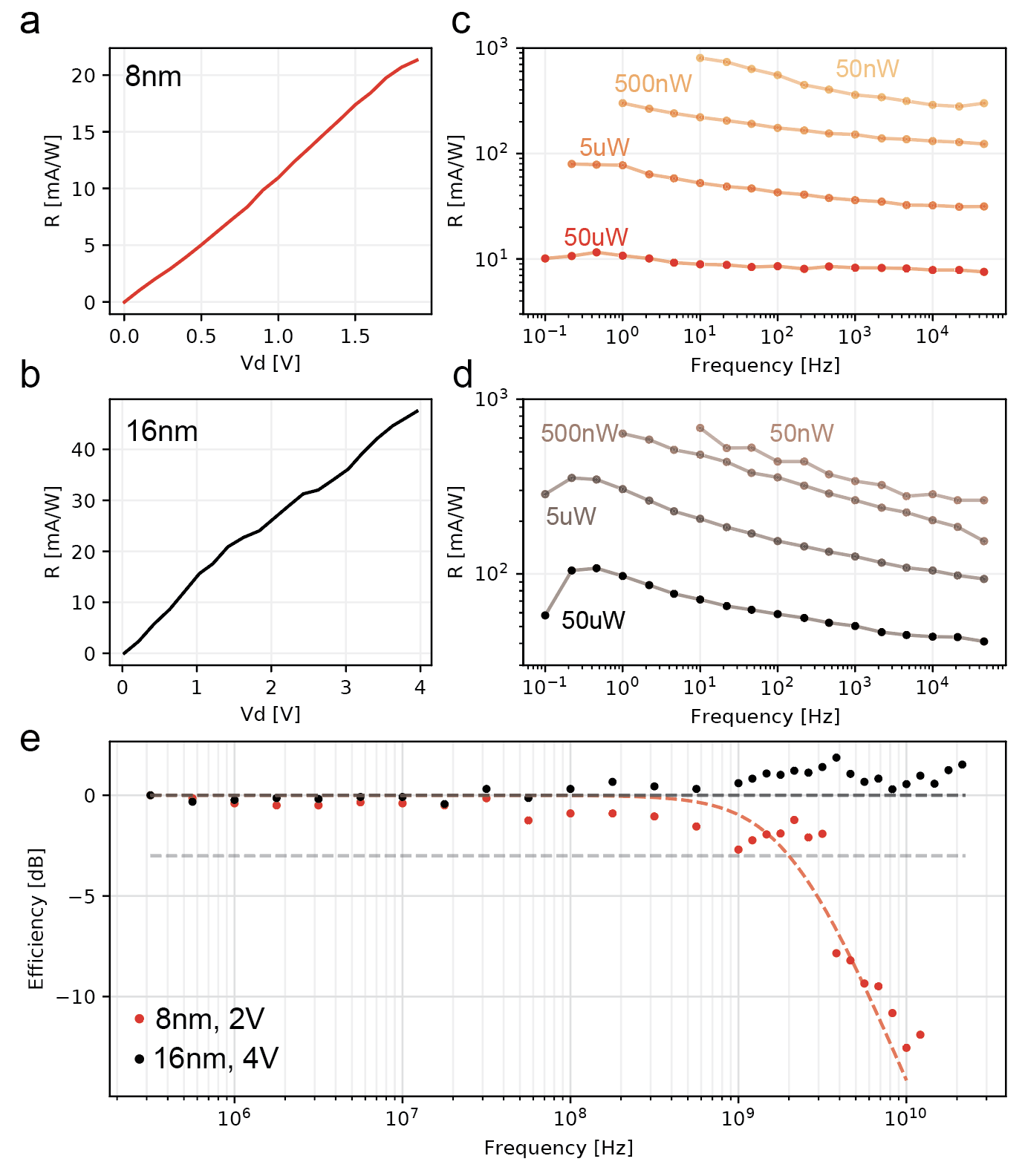}
\caption{\label{fig:Fig2}{\bf{Telecom Characterization}} \textbf{(a)} Voltage-dependent responsivity of a 100~$\mu$m wide 8~nm thick Te device measured at 50~$\mu$W optical power and 50 ~MHz. \textbf{(b)} Voltage dependent responsivity of a 20 $\mu$m wide 16~nm thick Te measured at 150~$\mu$W optical power and 10 ~kHz. \textbf{(c)} Low-power low-frequency characterization of a 5~$\mu$m wide 8~nm thick Te device. \textbf{(d)} Low-power low-frequency characterization of a 20~$\mu$m wide 16~nm thick Te device. \textbf{(e)} High speed measurements of 8 and 16~nm thick devices. 2 V and 4 V are used for 8 nm and 16 nm speed measurements. Efficiency normalized to 6~mAW$^{-1}$ for 8 nm devices and 11 mAW$^{-1}$ for 16 nm devices, measured with 150~$\mu$W input power. All measurements are at 1546~nm. The gray dotted line indicate -3 dB.}
\end{figure}

We next measure both 8 nm and 16 nm thick tellurium on silicon to study the effect of the carrier mobility. While 8 nm and 16 nm Te can both absorb 1550 nm light, the carrier mobility in 16 nm is significantly higher. While using thicker devices would increase mobility even further, this would also reduce the bandgap and allow signal obfuscation through the absorption of thermal photons. Hence, we investigate using Te thickness of 8 nm and 16 nm. We first verify the functionality of the devices by measuring the bias voltage dependent responsivity, shown in FIG 3a-b.
We observe that the responsivity increases at very low optical power and frequency as shown in FIG 3c and 3d. This is likely due to avalanching of carriers trapped at defect sites \cite{dan2018photoconductor}. As the carrier regeneration at these defects is a slow process, this limits the gain at higher input powers and at higher frequencies. We see a maximum responsivity of 0.8 AW$^{-1}$. 
At higher optical power levels and frequencies much larger than those at which we see gain, the responsivity is extremely uniform, as shown in FIG 3e, with 6 mAW$^{-1}$ at 8nm and 11 mAW$^{-1}$ at 16nm. We see that, due to the lower carrier mobility at the thinner Te thickness \cite{Zhao2021AM}, the frequency response of the 8nm devices roll off with a 3-dB bandwidth near 2 GHz while the 16nm devices show no degradation in responsivity out to at least 22 GHz, where our measurements were limited. From the roll-off frequency, we can estimate the transit time present in the device as defined by $\tau_{tr} = l^{2} / 2 \mu V_{d}$, where $\mu$ is the carrier mobility and $l$ is the separation length between the electrodes. For the Te photodetectors presented here, we use $l = 3 \times 10^{-6}$ m for ease of fabrication. From the measurements, we estimate mobility of 45 cm$^{2}$V$^{-1}$s$^{-1}$ for 8 nm Te, and mobility of more than 250 cm$^{2}$V$^{-1}$s$^{-1}$ for 16 nm Te, which are in good agreement with previous reports on the experimental measurements of tellurium with the corresponding thicknesses \cite{Zhao2021AM}. 
In addition, the waveguide integrated Te-photodetectors show excellent stability, as shown by measurements taken 1.5 months apart, Supplementary FIG S5. Within this time period, the Te-photodetectors were left in the air, and do not show noticeably different performance than immediately after the fabrication. This is consistent with a previous report \cite{tong2020stable} showing the air stability of Te flakes. 







\begin{figure*}[t]
\centering
\includegraphics[width=0.85\linewidth]{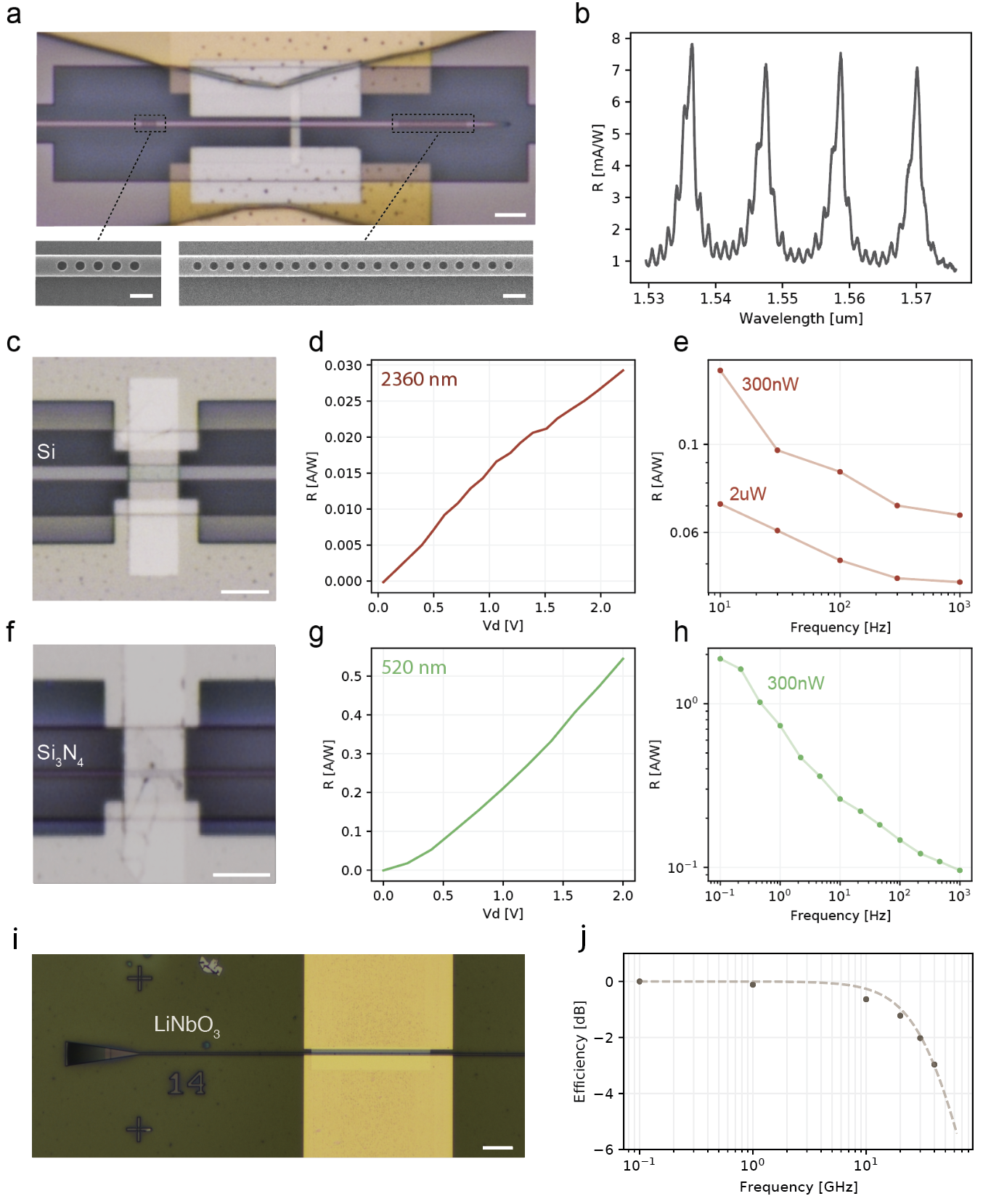}
\caption{\label{fig:Fig3}{\bf{Flexible Integration}} \textbf{(a)} Optical micrograph and SEMs of an integrated Fabry-Perot cavity that enhances a 1~$\mu$m wide 16~nm Te device absorption. Scale bars are 3~$\mu$m (top) and 500~nm (bottom). \textbf{(b)} Responsivity over wavelength of integrated Te-cavity device. Resonant enhancement shows more than half of the responsivity of a 20~$\mu$m long device with far less dark current. \textbf{(c)} Optical micrograph of a 16~nm thick device operating at 2360~nm (scale bar is 5~$\mu$m). \textbf{(d)} Voltage dependent responsivity measured at 2~$\mu$W input power and 1~kHz. \textbf{(e)} Low-power, low-frequency response of the same device. \textbf{(f)} Optical micrograph of a 16~nm thick device operating at 520~nm on a silicon nitride waveguide (scale bar is 5~$\mu$m). \textbf{(g)} Voltage dependant responsivity measured at 2~$\mu$W input power and 1~kHz. \textbf{(h)} Low-power, low-frequency response of the same device. \hl{ \textbf{(i)} Optical micrograph of thin-film Lithium Niobate on sapphire integration. Scale bar is 25~$\mu$m \textbf{(j)} Frequency dependant responsivity of LN device showing operation up to 40GHz. Measured with 100~$\mu$W input power at 1557~nm. Normalized to a responsivity of 17.5~mA/W. Dashed line shows a 3dB bandwidth of 40 GHz.}}
\end{figure*}

Moreover, as we can fabricate these detectors directly on top of existing photonic structures owning to the ease of integration, we can co-design the photonics and detectors in order to optimize their performance.
We demonstrate this co-design process by resonantly enhancing the absorption of a small-volume photodetector with an integrated Fabry-Perot cavity, shown in FIG 4a. 
The cavity is designed to critically couple the light from waveguide to the Te-photodetector for maximum absorption. This is achieved by matching the reflection of the input mirror to the calculated absorption of the small volume Te, which is then lithographically patterned on top.
As these detectors are operating in photoconductive mode, the dark current can add significant noise. Therefore, shrinking the active volume of the detector while maintaining a similar level of absorption increases the signal to noise ratio. To achieve this, we co-engineer the integrated photonic cavity and length of the deposited Te thin film that correspond to the absorption per an optical pass on the integrated cavity. Critically, the liftoff based definition process for Te thin film allows for precise control of the absorption per optical pass, and hence allows for better coupling to the integrated cavity.
As shown in FIG 4b, we are able to resonantly enhance the absorption of a 1~$\mu$m wide device, achieving a peak responsivity of 8 mAW$^{-1}$ compared to 11 mAW$^{-1}$ 
in a 20~$\mu$m device with a dark current of 25~$\mu$A compared to 500~$\mu$A. This shows more than an order of magnitude reduction with a comparable responsivity. \hl{We note that the quality factor of the resonant device is approximately 500, which corresponds to 3dB bandwidth of 370 GHz. Hence, this cavity integration would not compromise the device speed.}

To demonstrate the flexible integration and broadband operation of evaporated Te detectors, we integrate detectors onto silicon devices operating in the mid-infrared (FIG 4c-e) and onto silicon nitride devices operating in the visible range (FIG 4f-h). In both devices, we again see enhanced responsivities at low power and low frequency. The 2360nm mid IR silicon-based device achieves a maximum responsivity of 0.15 AW$^{-1}$.
The 520nm visible device with silicon nitride photonics achieves a responsivity of 1.9 AW$^{-1}$. 
This indicates that waveguide integrated Te-photodetectors can operate efficiently across a wide wavelength band and can operate with arbitrary photonic circuits underneath. 

\hl{Finally, we fabricate detectors on thin-film Lithium Niobate devices designed for 1550nm, FIG 4i. Using an external high speed electro-optic modulator, we measure the RF frequency dependant responsivity, FIG 4j. We find that the 3dB bandwidth is larger than but likely close to 40~GHz, limited by the RF probe. From this measurement we can get a more accurate estimation for the carrier mobility of 486~cm$^{2}$V$^{-1}$s$^{-1}$ (40 GHz bandwidth with 2.7~$\mu$m gap and 3V bias). Given this high mobility present in the tellurium films, by shrinking $l$ to $1 \times 10^{-6}$ m, which is still easily achieved with scalable photo-lithography, these photodetectors can potentially reach operation bandwidths close to 300 GHz. The RC bandwidth from the contact capacitance and resistance (measured contact resistance of 200~$\Omega$ and calculated 2~fF capacitance) is approximately 400 GHz, so the operation would not be RC bandwidth limited.} 

%
%
%
%
%

We demonstrated air-stable evaporated tellurium photodiodes that are easy to fabricate, can achieve high efficiency, and can be flexibly integrated onto numerous photonics platforms for wavelengths ranging from visible to mid infrared. Our thermal evaporation of Te thin films leaves the tellurium crystalline with extremely high carrier mobility, allowing us to demonstrate state-of-the-art detection bandwidths above 40 GHz, with beyond 100 GHz bandwidths easily achievable. \hl{These Te devices will functionalize the on-chip detection of light in many material systems that lacked integrated photodetection, opening up these platforms for novel applications and widespread deployment. For instance, the integration of high speed photodetectors on lithium niobate can enable single-chip photonic systems that take advantage of high speed electro-optic modulators \cite{wang2018integrated} for communication and frequency modulated continuous wave (FMCW) LiDAR. The integration of high efficiency photodetectors on flexible substrates can enable more complex and compact optical biosensors\cite{li2018monolithically}, and scalable integration of photodetectors operating at mid-IR can enable on-chip mid-IR spectrometer for environment sensing and pollution monitoring. Additionally, evaporated Te has shown promise for high performance field-effect transistors, which in combination with their photo-detectivity opens up a large space for photonic-electronic co-integration for complex systems beyond silicon photonics.} \\

\bibliography{Reference}

\clearpage 

\noindent{\bf Methods}\\
\\
\noindent\textbf{Device Fabrication}
The photonic circuits in silicon and silicon nitride were patterned with electron-beam lithography using JEOL EBX6300-FS electron lithography system with ZEP 520A as the mask. The patterned waveguide substrates were etched using reactive ion etch (RIE), and cleaned piranha solution. Diluted H-SiOx (HSQ) is spin-coated and annealed at 500 °C.
The electrode pads for the Te were patterned by photolithography using an i-line photoresist and mask aligner (Karl Suss), followed by thermal evaporation of 5 nm/25 nm titanium/gold and lift-off process. Channel regions were defined by e-beam lithography using PMMA (C4, Microchem,). Te thin films were subsequently deposited with a thermal evaporation system (E306A, Edwards), where the substrate temperature was decreased to -80 °C at a base pressure of $2 \times 10^{-6}$ mTorr and the deposition rate was maintained to be 10 Ås$^{-1}$ [28]. The final electrode contacts were patterned by e-beam lithography with a lower beam current to prevent electron beam damage to the Te, followed by thermal evaporation of 30 nm palladium and lift-off.
\\

\noindent\textbf{Low Frequency Measurements} To measure at low frequency, we put the devices in series with a 1k$\Omega$ resistor and apply a bias across the device and resistor. For the telecom measurements we use an electro-optic modulator (Optilab) to modulate the driving laser (Toptica) and a lock-in amplifier (Stanford Instruments) to read the AC current through the resistor generated by the device. For mid-IR measurements, we use an OPO laser (ARGOS) modulated by a free space chopper, and for visible measurements, we use a directly modulated green laser diode (Thorlabs). 
\\

\noindent\textbf{High Frequency Measurements} To measure at high frequency, we bias the devices with an RF probe (Cascade Mircotech) through a high frequency bias-T (Picosecond). We drive the devices optically with a laser (Toptica) modulated by an EOM (Optilab). The RF modulation signal is generated by an AWG (Keysight) and the device current is read out with a spectrum analyzer (Agilent). A schematic of the measurement setup is shown in the Supplementary FIG S6.
\\

\noindent\textbf{Acknowledgments}
\noindent The authors thank Melissa Guidry and Kasper Van Gasse for assistance with the experimental setup, and Rahul Trivedi and David A.B. Miller for insightful discussions. G.H.A. acknowledges support from STMicroelectronics Stanford Graduate Fellowship (SGF) and Kwanjeong Educational Foundation. A.W. acknowledges the Herb and Jane Dwight Stanford Graduate Fellowship (SGF) and the NTT Research Fellowship. H.K. acknowledges support from a Samsung Scholarship. J.F.H. acknowledges support from the National Science Foundation Graduate Research Fellowship Program (Grant No. DGE-1656518). Authors from Stanford acknowledge funding support from DARPA under the LUMOS program. The work at Stanford was supported by the Department of Energy under FWP 100786, Atoms-to-Systems Co-Design: Transforming Data Flow to Accelerate Scientific Discovery project at SLAC National Accelerator Laboratory, under contract DE-AC02-76SF00515. The work at Berkeley was funded by the U.S. Department of Energy, Office of Science, Office of Basic Energy Sciences, Materials Sciences and Engineering Division under contract no. DE-AC02-05-CH11231 (EMAT program KC1201). \\

\clearpage

\renewcommand{\thefigure}{S\arabic{figure}}
\setcounter{figure}{0}

\begin{figure*}[h]
\centering
\includegraphics[width=0.7\linewidth]{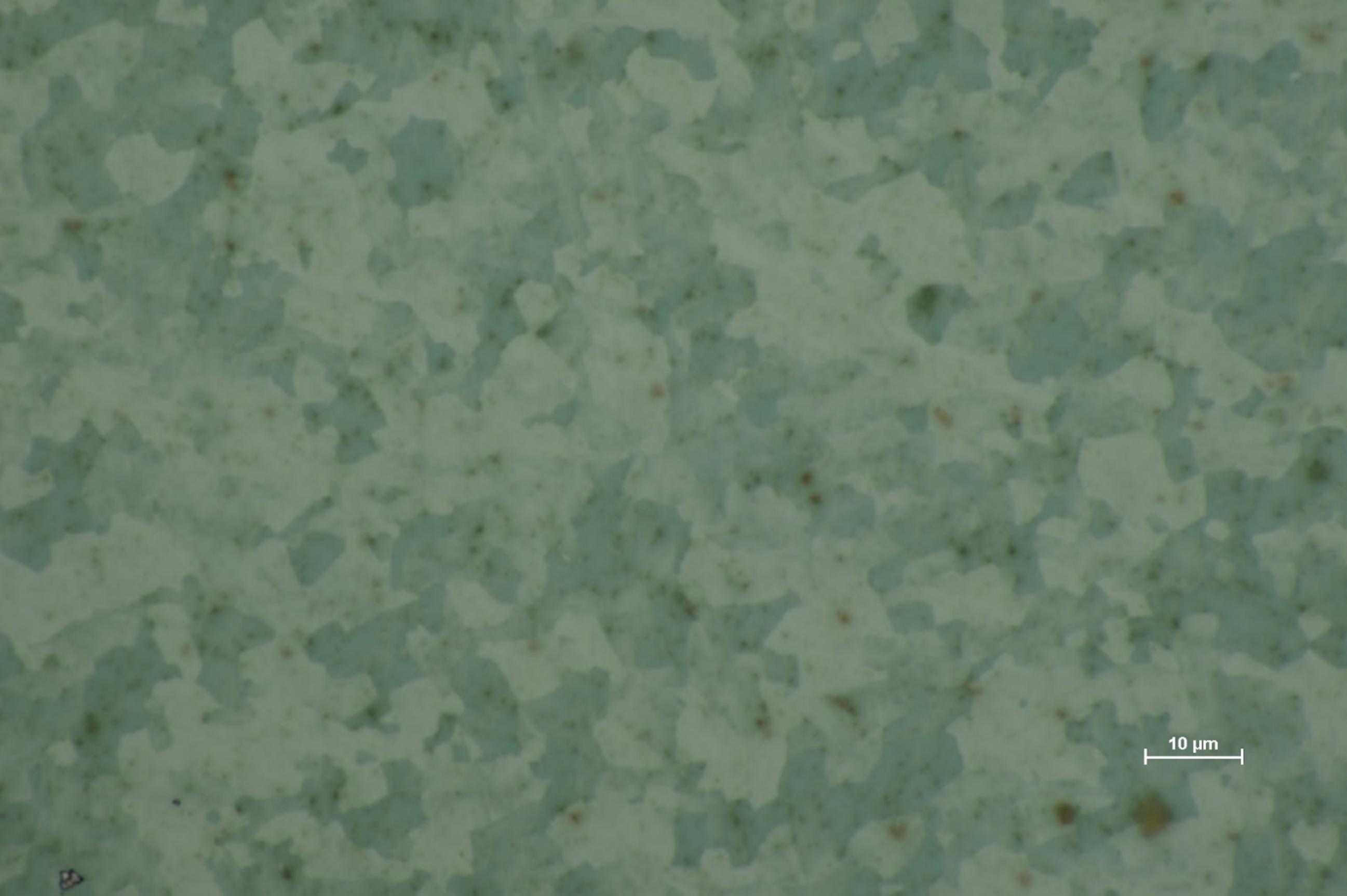}
\caption{\label{fig:Fig1}{\bf{As deposited Te thin films}} Optical image of as-evaporated Te thin films on silicon substrate. The image shows clear poly crystalline grains.}
\end{figure*}

\begin{figure*}[h]
\centering
\includegraphics[width=0.7\linewidth]{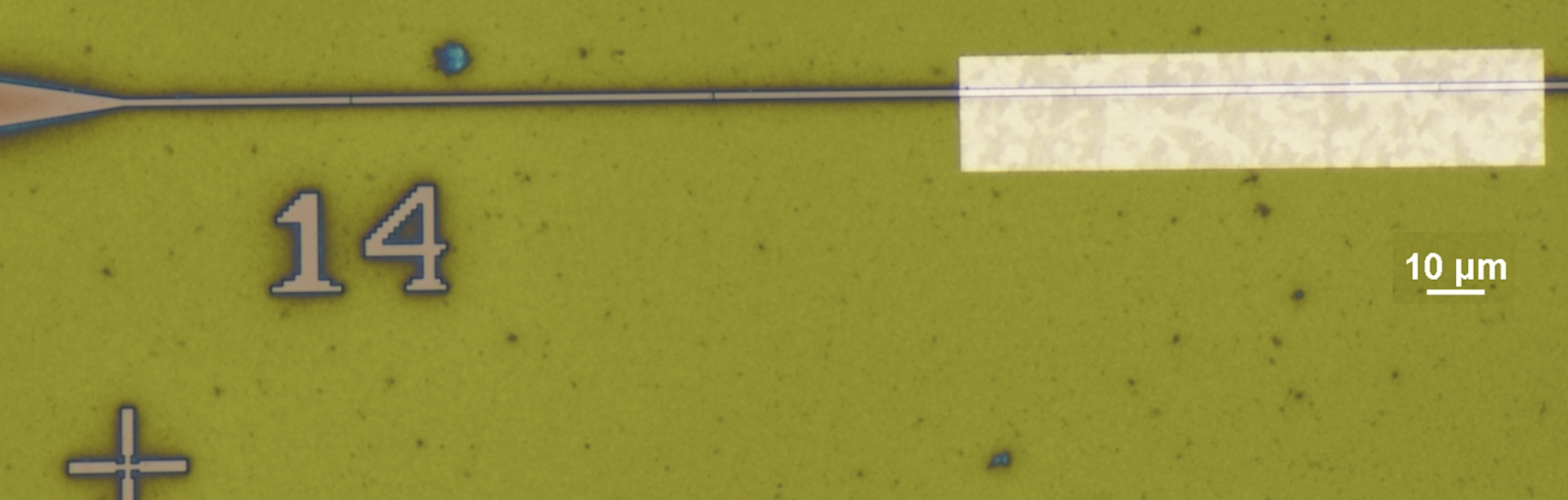}
\caption{\label{fig:Fig1}{\bf{As-deposited, post-liftoff Te thin films}} Optical image of post lithographic definition of Te thin films on lithium niobate PIC}
\end{figure*}

\begin{figure*}[h]
\centering
\includegraphics[width=0.7\linewidth]{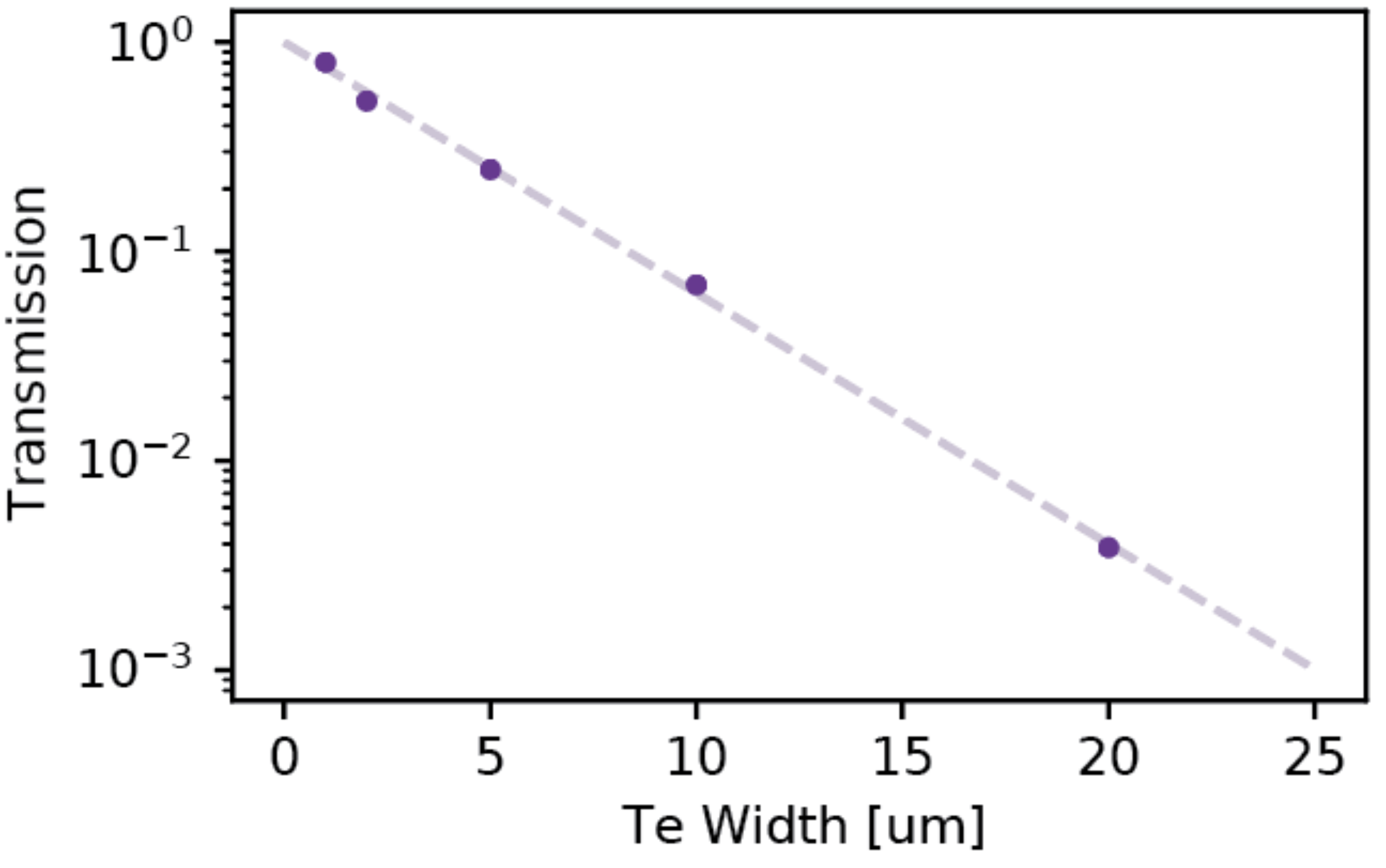}
\caption{\label{fig:Fig1}{\bf{Optical transmission measurement through integrated Te thin film}} Transmission through 16 nm Te devices of varying widths of Te thin films. 10  \textmu m devices show over 90\% absorption and 20 \textmu m devices show over 99\% absorption. Fit corresponds to an absorption coefficient of 2760 cm$^{-1}$. }
\end{figure*}

\begin{figure*}[h]
\centering
\includegraphics[width=0.7\linewidth]{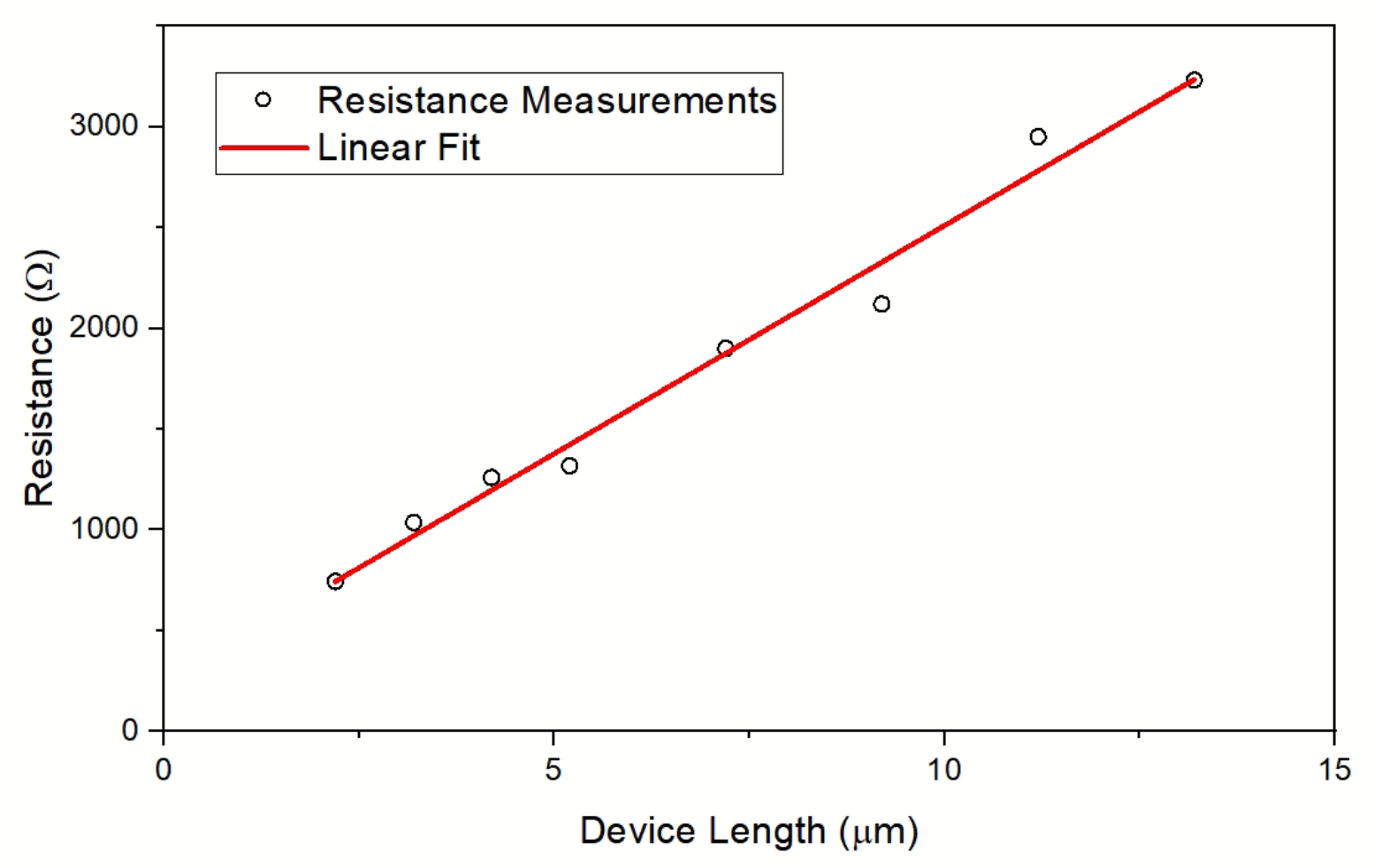}
\caption{\label{fig:Fig1}{\bf{Transfer Length Measurement of Te thin films with palladium contacts}} Transfer Length Measurement (TLM) of evaporated Te films resistance with Pd contacts, which are used for all the fabricated Te photodetectors. The y-intercept of the fit indicates approxmiate contact resistance of 247 Ohms.}
\end{figure*}

\begin{figure*}[h]
\centering
\includegraphics[width=0.7\linewidth]{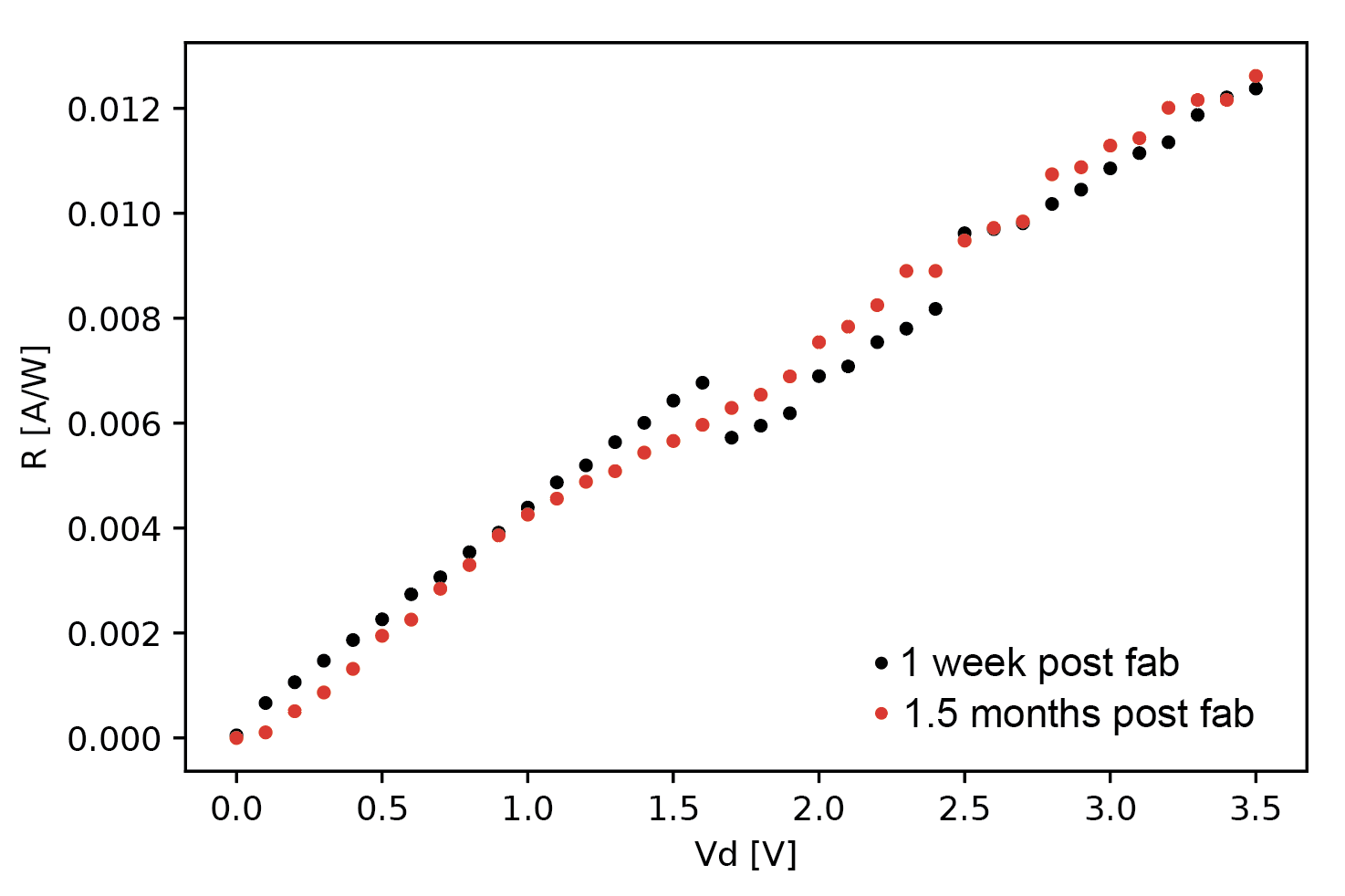}
\caption{\label{fig:Fig1}{\bf{Environmental Stability}} \textbf{(a)} Drain Voltage (Vd) dependent 1 6nm thick device responsivity measured one week and one and half months post fabrication. Measured at 178MHz with 150 \textmu W input power.}
\end{figure*}

\begin{figure*}[h]
\centering
\includegraphics[width=0.7\linewidth]{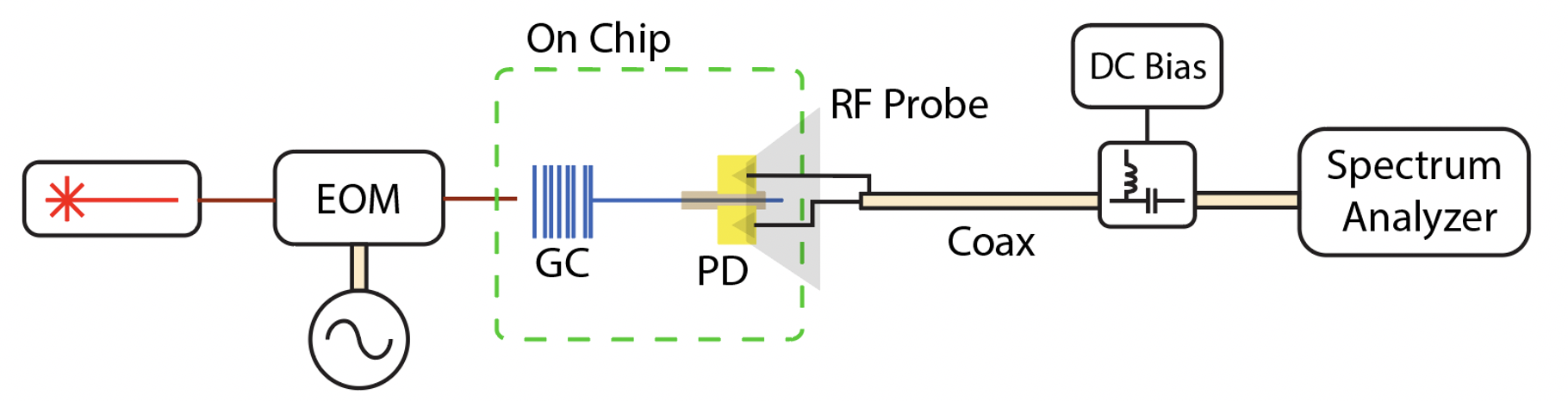}
\caption{\label{fig:Fig1}{\bf{High Frequency Measurement Schematic}} \textbf{(a)} Chip is driven optically with a laser modulated by an EOM. The tellurium device is biased through a bias tee and the current generated is read out by a spectrum analyzer. }
\end{figure*}

\end{document}